# Spectroscopic Properties of Nanotube-chromophore Hybrids

*Changshui Huang[†], Randy K. Wang[‡], Bryan M. Wong[⊥], David McGee[£], François Léonard[⊥], Yunjun Kim[†], Kirsten F. Johnson[†], Michael S. Arnold[†], Mark A. Eriksson[‡,*], and Padma Gopalan[†,§,*]*

[†] Department of Materials Science & Engineering, University of Wisconsin-Madison, 1150 University Ave., Madison, Wisconsin 53706

[‡] Department of Physics, University of Wisconsin-Madison, 1150 University Ave., Madison, Wisconsin 53706

[§] Department of Chemistry, University of Wisconsin-Madison, 1150 University Ave., Madison, Wisconsin 53706

[⊥] Sandia National Laboratories, Livermore, California 94551

[£] Department of Physics, Drew University, 36 Madison Ave., Madison, New Jersey 07940

[*] E-mail: pgopalan@cae.wisc.edu, maeriksson@wisc.edu

**ABSTRACT**

Recently, individual single-walled carbon nanotubes (SWNTs) functionalized with azo-benzene chromophores were shown to form a new class of hybrid nanomaterials for optoelectronics applications. Here we use a number of experimental techniques and theory to understand the binding, orientation, and nature of coupling between chromophores and the nanotubes, all of which are of relevance to future optimization of these hybrid materials. We find that the binding energy between chromophores and nanotubes depends strongly on the type of tether that is used to bind the chromophores to the nanotubes, with pyrene tethers resulting in more than 90% of the bound chromophores during processing. DFT calculations show that the binding energy of the chromophores to the nanotubes is maximized for chromophores parallel to the nanotube sidewall, even with the use of tethers; second harmonic generation shows that there is nonetheless a partial radial orientation of the chromophores on the nanotubes. We find weak electronic coupling between the chromophores and the SWNTs, consistent with non-covalent binding. The chromophore-nanotube coupling, while weak, is sufficient to quench the chromophore fluorescence. Stern-Volmer plots are non-linear, which supports a combination of static and dynamic quenching processes. The chromophore orientation is an important variable for chromophore-nanotube phototransistors, and our experiments suggest the possibility for further optimizing this orientational degree of freedom.

**INTRODUCTION**

Light-triggered changes in biological molecules, which enable various functions such as vision,[1] photosynthesis, and heliotropism,[2] have long inspired materials chemists to mimic these phenomena to create new synthetic materials and devices. One example is the molecule retinal undergoing a *cis-trans* isomerization in response to light,[3] creating a cascade of events leading to visual recognition.[4] Synthetic versions of retinal include switchable stilbene, or azo-benzene containing molecules. Azo-benzenes undergo a reversible photoisomerization from a thermally stable *trans* configuration to a meta-stable *cis* form. Dipolar chromophores based on azo-benzene structure are photochemically stable, can be reversibly switched $10^5$ to $10^6$ times before bleaching, and can be chemically tuned at the donor and acceptor end to alter the magnitude of the dipole moment.[5] The *cis-trans* isomerization of a range of chromophores has been studied extensively in solution and as monolayers on gold-coated[6,7] flat substrates or on silicon substrates using scanning tunneling microscopy (STM).[8] Hence, azo-benzene chromophores constitute a well-understood switching unit for attachment to nanotubes to create new hybrid materials.[9] The reversible, wavelength-selective isomerization and the accompanying conformational change provide an important handle for optical modulation of electrical and electro-optic properties of nanotubes. We recently demonstrated an optically active nanotube-hybrid material by non-covalent functionalization of SWNT field-effect transistors with an azo-based chromophore.[10] Upon UV illumination, the chromophore undergoes a *trans-cis* isomerization leading to charge redistribution near the nanotube. The resulting change in the local electrostatic environment leads to a shift in the threshold voltage and increased conductivity of the nanotube.[11] The functionalized transistors showed repeatable switching for many cycles, and the low (100 $\mu W/cm^2$) intensities necessary to optically modulate the transistor are in stark contrast to measurements of intrinsic nanotube photoconductivity, which typically require 1 kW/cm$^2$ intensity.[12] More recently this approach was used to demonstrate photodetection with tunability over the visible range and using covalent functionalization on multiwalled nanotubes.[13-15] In order to improve the efficiency, stability, and lifetime of chromophore-functionalized SWNT devices, it is important to understand the chromophore/SWNT interactions, including the role of tethers in binding the chromophores to the SWNTs.

In this paper, we report the spectroscopic analysis of SWNT-azo-benzene chromophore hybrid systems using complementary experimental techniques. For the chromophore, we use Disperse Red 1 (DR1), a well studied and commercially available azobenzene chromophore (pseudo stilbene type azo compound), and we study a series of three increasing tether strengths: (1) unmodified DR1 with no added tether (DR1U), (2) anthracene-functionalized DR1 (DR1A), and (3) pyrene-functionalized DR1 (DR1P) (see Scheme 1). The first part of our studies focus on thin-films, show that the binding energy and hence the surface coverage of the chromophores on the nanotubes is strongly influenced by the strength of the tether and we address the question of the orientation of the molecules on the nanotubes by SHG measurements. These studies have direct relevance to the sensitivity of the phototransistors. The second part of our studies focus on the solution characterization of the hybrids to evaluate the nature of the electronic coupling. We find weak electronic coupling between the chromophores and the SWNTs, consistent with non-covalent binding, as verified by *ab initio* calculations. Our experiments show strong fluorescence quenching of the chromophores upon binding to the SWNTs. The mechanistic insight gained from these solution studies aid in optimization of the chromophore/SWNT system by rational design of the chromophore structure.

We present detailed spectroscopic characterization of these hybrid systems by fluorescence, Raman, UV-Vis, and X-ray photoelectron spectroscopy (XPS). Theoretical results from Density Functional Theory (DFT) calculations provide quantitative information on the binding energy between the chromophores and SWNTs as well as the electronic coupling. While DFT calculations indicate parallel binding is preferred, second harmonic generation (SHG) characterization indicates a nonzero net radial orientation of chromophore dipoles perpendicular to the SWNTs. We examine the mechanism of fluorescence quenching and the effect of increasing binding strength on the stability of the hybrids.

**RESULTS AND DISSCUSSION**

**Evaluation of the strength of binding:**

SWNTs can interact with a DR1 unit by π–π stacking interactions with the two benzene rings, and/or by π–π stacking interactions with anthracene and pyrene tethers, if present.[22-24] However, the strength of binding and the resulting surface coverage can vary greatly, and both are relevant for the fabrication of useful devices.

Raman spectroscopy was used to characterize the SWNT-chromophore hybrids, as it allows the measurement of vibrational modes of SWNTs.[25] Raman spectra were collected in the 1200 to 1800 cm$^{-1}$ range following excitation with a 532 nm laser (Figure 1A-D). Raman spectra of SWNTs typically consist of a graphitic or G-band[25] from highly ordered sidewalls, while disorder in the sidewall structure results in the D-band. From the Raman spectra (orange line) of pristine SWNTs, the D-band is observed in the region of 1300 – 1350 cm$^{-1}$; and the G-band in the 1500 – 1650 cm$^{-1}$ range. The asymmetric shoulder at 1540 cm$^{-1}$ on the G-band is due to electron-phonon coupling in bundles of SWNTs.[26] The Raman spectra of the hybrids show the characteristic peaks from DR1 in the 1200 – 1600 cm$^{-1}$ range (See supporting information Figure 2s).[27,28] When compared to the pristine SWNTs, the hybrids show the following changes: 1) an increased peak intensity ~ 1330 cm$^{-1}$, due to contributions from the N–O and C–N stretching vibrations of the chromophore, 2) the appearance of new peaks around 1350 – 1500 cm$^{-1}$ attributed to N=N, $C_{ph}$–$N_{Me}$, C–C stretching vibrations, and C–C, C–N in-plane bending vibrations of the chromophore,[29,30] and 3) increased intensity of the peak at 1500 – 1550 cm$^{-1}$ due to contributions from the C–C stretching vibrations from the chromophores. Hence, quantitative comparison of the peak intensities in the 1200 to 1800 cm$^{-1}$ range can give us an estimate of the amount of bound chromophore. Upon washing the DR1U/SWNT (Figure 1A) hybrid with methanol to remove any weakly bound molecules, the intensity of the characteristic peaks from the chromophores (region $A_1$ in Figure 1) decreased significantly (region $A_2$ in Figure 1). In contrast, the peaks in the DR1P/SWNT decreased only marginally, as shown in Figure 1C. The information from Figure 1A-C is summarized in Figure 1D as a plot of the relative binding efficiencies of the three types of bound molecules to nanotubes. Comparison of the normalized area before and after methanol washing, shows retention of 30%, 40% and 60% of DR1U, DR1A and DR1P respectively, hence confirming that pyrene tether provides a more stable structure compared to stacking by benzene rings or anthracene units. As shown in the supporting information, similar results were obtained by comparing the UV-vis spectrum of the nanohybrids.

X-ray photoelectron spectroscopy (XPS) provides another means to probe the surface chemical composition and to evaluate the surface coverage. Figure 2A shows the N (1s) XPS spectra of nanohybrid films. After washing with methanol, the decrease in the N (1s) peak intensity was in the order DR1U/SWNTs > DR1A/SWNTs > DR1P/SWNTs. The following

bonds were assigned (Figure 2B-C): N=N (397.38–401.53 eV), N–C (398.28–403.53 eV), $NO_2$ (403.18–408.29 eV); $sp^2$ C=C and $sp^3$ C–C (282.38–286.21 eV), C–O (283.91–287.19 eV), C–N (283–287.5 eV), >C=O (285.21–287.98 eV), –COO and O–COO (286.52–289.09 eV), $\pi$-$\pi$* transitions associated with phenyl rings (288.51–291.1 eV).[31] The N (1s) peak around 400.3 eV has contributions from the azo, nitro, and the amine groups on the chromophore. The C (1s) peak centered at 284 eV, has contributions from both the carbon atoms in the chromophore and the SWNTs. The ratio of nitrogen atoms to nanotube carbon atoms was calculated from the XPS spectra (Figure 2D). The calculated coverage's are 0.64, 0.67, and 3.12 molecules per 100 nanotube carbon atoms for DR1U/SWNTs, DR1A/SWNTs, and DR1P/SWNTs hybrids respectively. After washing with methanol, the differences between these three systems are dramatic. The surface coverage of DR1U and DR1A on nanotubes decreased by 78% and 76% respectively, whereas that of DR1P decreased by only 10%. These results are in agreement with the Raman and UV-Vis data discussed earlier, and they confirm that the pyrene tether plays an important role in forming stable nanotube hybrids.

These experimental results can be compared with theoretical predictions of binding energies and electronic structures of the various chromophore-SWNT hybrids, using all-electron *ab initio* calculations.[32,33] In these calculations, the binding energy are affected by the geometrical orientation of the chromophore on the nanotube. Figure 3 shows the relaxed atomic structures and electronic band structures for three configurations: (a) DR1U, (b) DR1P perpendicular to the SWNT axis, and (c) DR1P parallel to the SWNT axis. These calculations indicate that DR1P in the parallel configuration (1.55 eV) shows stronger binding interaction to SWNTs than DR1U (0.79 eV), while the perpendicular configuration of DR1P (0.64 eV) has smaller binding energy as DR1U. (It is worth noting that the binding energy of DR1P in the parallel configuration is roughly equal to the sum of the energies of DR1U and of DR1P in the perpendicular orientation.) According to the experimental UV spectra, Raman spectra and XPS, DR1P shows stronger interaction to the nanotubes than DR1U, which suggests that at least some parallel orientation of DR1 is present in the experimental samples. If the binding of chromophores to nanotubes were entirely parallel to the surface it would require reexamining the mechanism of gating of phototransistors,[10,14] which was explained based on a perpendicular orientation of the chromophores on the SWNTs. Such a parallel orientation would result in a macroscopically symmetric system with no net orientation of dipoles, so that no change would be observed in electrostatic potential around the nanotubes when the hybrids are exposed to light. To address this issue, we performed optical second harmonic generation experiments on the nanotube hybrids both in the as-cast form and upon UV-illumination.

Second harmonic generation (SHG) has proven to be a sensitive probe for the determination of acentric molecular ordering in monolayers and thin film.[34,35] It should be noted that SHG experiment have not been employed so far in the analysis of chromophore functionalized nanotubes, and our results as explained below show that these measurements can infact provide unique information about the orientation of these molecules. When irradiated by a laser pulse, a collection of weakly interacting chromophores will generate a second-harmonic pulse, with an intensity that depends on the molecular second order hyperpolarizability $\beta$ and a macroscopic order parameter describing the average orientation of a chromophore with respect to an axis of symmetry (in this case, normal to the SWNT film). Since isomerization of DR1 results in an approximate 5-fold decrease in $\beta$ in going from the *trans* to the *cis* conformation, monitoring SHG emission from the SWNT-chromophore hybrid during UV exposure would provide insight on the net chromophore orientation[36-38]. We observed that the DR1P/SWNT film exhibited SHG immediately following fabrication, indicating a degree of chromophore alignment perpendicular to SWNTs. This was not observed for pristine SWNTs. To confirm the chromophore role in generation of the SHG signal, further measurements were performed while irradiating the DR1P/SWNT films for 1 minute with 365 nm UV light, followed by 1 minute without UV light (Figure 4). The DR1P/SWNT film shows a clear dependence of the SHG signal on 365 nm UV light. Upon illumination, the SHG drops noticeably, taking 2-4 seconds to reach steady state. Following removal of the UV, the SHG returns to its initial value. These results can be repeated for several cycles. The change that occurred during this process was consistent with the isomerization of DR1P oriented in the perpendicular direction. An AmineP/SWNTs (Scheme 1) film was prepared as a control as it does not have the switchable azo group. No SHG or dependence on the UV light was observed, confirming the importance of the isomerization of DR1P. Collectively, these results suggest that the adsorption of chromophores on the nanotubes is heterogeneous: in addition to the parallel configuration of DR1P, there are also perpendicular configurations on the SWNTs.

In all three systems, however, the electronic properties near the bandgap in the nanotubes are not strongly affected. As shown in Figure 3, the valence and conduction bands of the nanotube shift, but they do so together, with very little change in the bandgap. Further, the chromophore energy levels near the HOMO and LUMO do not hybridize with the nanotube band structure. Moving above the LUMO or below the HOMO a small degree of hybridization between the CNT and chromophore levels is seen. Overall, the DFT calculations indicate that there is weak but nonzero electronic coupling between the chromophore and the nanotube, consistent with the small shifts observed in the G band in the Raman spectra (Figure 3S in the supplementary).

**Fluorescence Quenching upon Binding between Chromophores and Nanotubes:**

The three chromophore samples showed a strong fluorescence emission when not bound to nanotubes (Figure 5). However, in the presence of nanotubes, the fluorescence intensity (F) drops dramatically compared to the initial fluorescence intensity ($F_0$). For a given concentration of chromophores, as the concentration of nanotubes increases, the emission from the chromophores decreases. The efficiency of quenching($1-F/F_0$) for DR1U/SWNTs, DR1A/SWNTs, and DR1P/SWNTs are 84.2%, 92.0%, and 95.6%, respectively. From the Stern-Volmer plots ($F_0/F$ as a function of nanotube concentration), in all three cases a non-linear plot with an upward curvature was observed. The non-linear plot is typically associated with a combination of both static and dynamic quenching i.e., by both collisions and by complex formation with the chromophores. The absorption spectrums of the chromophores with and without SWNTs are presented in Figure 5A-C. The absorption maxima of DR1U, DR1A, and DR1P show red-shifts of 5 to 10 nm upon addition of the SWNTs. The red shift is attributed to changes in energy levels upon binding of the chromophores to the nanotubes, and is associated with the static component of quenching mechanism. The interactions between the chromophores and the nanotubes is also supported by the slight upfield shift in the G band in Raman, upon functionalization with all three chromophores (see Figure 3S in the supporting information).[36-38]

Two main modes for quenching of the photo-excited fluorophores by CNTs have been discussed in the literature: energy transfer (i.e., Forster resonant energy transfer) or electron transfer (i.e., photoinduced electron transfer).[39-41] We measured the PL emission from the nanotubes pumped by absorption of the chromophores (Figure 4S in supplementary information). The excitation source was 490nm (close to the $\lambda_{max}$ of the chromophore), and the chromophore concentration was 5 μM. In all three cases no detectable enhancement in the band-gap PL of nanotubes was observed, which points to lack of significant energy transfer between the chromophores and the nanotubes. This leaves the likely mechanism to be charge transfer.[39,42-44] The DFT calculations discussed above do not show significant charge transfer in the ground state; thus, if charge transfer is present it must occur in an excited state. It should be noted that we cannot completely rule out energy transfer given the presence of significant amount of metallic tubes in the HIPCO sample.

**CONCLUSIONS**

Three azo-benzene based chromophores (DR1U, DR1A, and DR1P) were used to study the interaction of chromophores with SWNTs and the effect of the tethers, if present. These experiments highlight the importance of having a strong $\pi$-$\pi$ stacking group such as pyrene in creating stable nanotube hybrids, while preserving the electronic structure of the nanotubes. Our main conclusions are: (a) UV-Vis, Raman and XPS indicate that pyrene forms the strongest tether to nanotubes compared to anthracene or unmodified molecules; (b) DFT calculations show that the binding energy of DR1P with parallel orientation to the nanotube has the largest binding energy; (c) SHG measurements applied for the first time to chromophore functionalized nanotubes support the presence of perpendicular component to the overall orientation of DR1P on the SWNTs, suggesting heterogeneous adsorption of the functionalized chromophore; (d) the fluorescence of the dipolar chromophores is quenched upon binding to nanotubes; (e) Stern-Volmer plots are non-linear, which supports a combination of static and dynamic quenching processes; and (f) PL measurements show no significant energy transfer between the chromophores and the nanotubes leaving charge transfer as the predominant mechanism. The DFT calculations show that if charge transfer is present it must occur in an excited state. This is relevant for the photogating of the choromophore/SWNT transistors,[10,14] as it motivates our future work on both tailoring the dipole moment as well as improving the optical activity of the nanotube-hybrid material by selecting specific chiral distributions.

**METHODS**

**Chromophore/SWNT solutions:** 1 mg HiPCO grown SWNTs (Unidym, raw powder) were homogenized in 10 mL of O-dichlorbenzene (ODCB) from Sigma-Aldrich in a bath sonicator for 1 hr and sonicated by Fisher Scientific Model 500 Sonic Dismembrator with horn-tip for 10 min under ambient conditions. ODCB was chosen because nanotubes are highly dispersible in this solvent.[16] These mixtures were subjected to centrifugation (11000 rpm) for 1.5 hr to remove bundles and catalyst residues. The supernatant was centrifuged for an additional 2 hrs and used as the stock dispersion assuming that the final SWNTs concentration is 1X. Chromophores were added to the SWNT suspensions to obtain a concentration of 0.5 mM. Unmodified Disperse Red 1 (DR1U) (Sigma-Aldrich) was used as received. DR1P and DR1A were synthesized based on a previous literature report.[10]

**Chromophore/SWNT films:** 2 mL of SWNT-chromophore solution were drop cast onto glass slides (3 inch ×1 inch ×1mm), dried for 24 hr under ambient conditions, and followed by drying in vacuum for 48 hr at room temperature. To remove the excess unbound chromophores the prepared SWNT-chromophore films were dipped into 10 ml methanol for 30s and dried under ambient conditions.

**Characterization:** UV-Vis spectra of the hybrids on the glass slides were recorded with a Varian Cary 50 Bio UV-Visible spectrophotometer and Jasco 570 UV-Vis-near IR spectrometer. Fluorescence spectra were taken with an ISS PC1 photon counting fluorometer (ISS Instrument Inc., Champaign, IL) using 490 nm excitation. Raman spectra of the hybrids on the glass slides were measured using a Horiba Jobin Yvon LabRAM ARAMIS Raman Confocal microscope with 532 nm diode laser excitation.

XPS of the hybrids on the glass slides were measured with a Perkin Elmer 5400 ESCA spectrometer under Mg Kα x-ray emission. The surface coverages of the chromophores on SWNTs were calculated using the method described below.[17,18] The overall number of the nitrogen atoms per unit area, for example, can be calculated using:

$$N(N) = \frac{A_N}{S_N} \frac{S_{Si}}{A_{Si}} (\rho_{Si,SiO_2} \lambda_{Si,SiO_2} \sin(\theta)) \exp^{t/\lambda_{N,organic}\sin(\theta)} / \exp^{t/\lambda_{Si,organic}\sin(\theta)}$$

The number of carbon atoms can be calculated similarly using :

$$N(C) = \frac{A_C}{S_C} \frac{S_{Si}}{A_{Si}} (\rho_{Si,SiO_2} \lambda_{Si,SiO_2} \sin(\theta)) \exp^{t/\lambda_{C,organic}\sin(\theta)} / \exp^{t/\lambda_{Si,organic}\sin(\theta)}$$

Where, $N(N)$ is the number of N atoms per unit area.

$N(C)$ is the number of C atoms per unit area.

$A_C$, $A_N$ and $A_{Si}$ are the integrated XPS peak areas for C, N and Si peaks, respectively.

$S_{Si}$ and $S_N$ are the sensitivity factors of N and Si (including effects of the XPS asymmetry parameter and energy-dependent analyzer transmission),[19] respectively.

$\rho_{Si,SiO_2}$ is the number of Si atoms per unit volume in glass.

$\lambda_{Si,SiO_2}$ is the inelastic mean free path (IMFP) of Si photoelectrons in $SiO_2$.

*t* is the thickness of the layer.

$\lambda_{N,organic}$, $\lambda_{C,organic}$, and $\lambda_{Si,organic}$ are the IMFP of N, C and Si respectively in the organic self-assembled monolayer films.

The attenuation lengths of self-assembled monolayer was fit by the empirical equation $\lambda(\text{Å}) = 9.0 + 0.022E(\text{eV})$, where $E(\text{eV})$ is the kinetic energy in electron volts,[20] yielding $\lambda_{N,organic}$=2.8 nm, $\lambda_{C,organic}$ =3.1 nm, and $\lambda_{Si,organic} \approx$3.3 nm. The angle θ is the take-off angle of photoelectrons with respect to the sample plane ($\theta=45°$). The ratio of the nitrogen atoms and carbon atoms was obtained by dividing the equation 1 by 2:

$$\frac{N(N)}{N(C)} = \frac{A_N}{A_C} \frac{S_C}{S_N} exp^{t/\lambda_{N,organic}\sin(\theta)} / exp^{t/\lambda_{C,organic}\sin(\theta)}$$

Since $\lambda_{N,organic} \approx \lambda_{C,organic}$ and $t < \lambda$ for thin organic layer, we can conclude that $exp^{t/\lambda_{N,organic}\sin(\theta)} / exp^{t/\lambda_{C,organic}\sin(\theta)} \approx 1$. The ratio of nitrogen to carbon atoms is therefore approximated to $\frac{N(N)}{N(C)} = \frac{A_N}{A_C} \frac{S_C}{S_N}$. Chromophore contributes to both the N and the C peaks whereas the SWNTs contribute only to the carbon peak. From the molecular structure of the chromophore the ratio of N to C atoms is fixed at 4:16, 4:31, and 4:35 for DR1U, DR1A and DR1P respectively. Hence, we can calculate the coverages in terms of molecules per 100 carbon atoms for each system.

Second harmonic generation experiments were conducted with a Q-switched 1064 nm pulsed Nd:YAG laser with a 50 mJ pulse energy and 10 ns pulse width. Frequency-doubled light at 532 nm was detected with a photomultiplier and gated boxcar integrating electronics.

**DFT calculations:**

Density functional theory (DFT) calculations were performed using the recent M06-L functional, which is designed specifically for noncovalent and $\pi$-$\pi$ stacking interactions.[21] Geometry optimizations for all three chromophores were obtained using the M06-L functional in conjunction with an all-electron 3-21G gaussian basis set. At the optimized geometries, single-point energies with a larger 6-31G(d, p) basis set were used to calculate final binding energies. To provide further insight into electronic properties, we also computed the electronic band structure at the same M06-L/6-31G(d, p) level of theory for all the SWNT-chromophore hybrids. For all three chromophores, a (10, 0) semiconducting SWNT was chosen as a representative model system, and calculations were performed using a one-dimensional supercell along the axis of the nanotube. Since the chromophore molecules are over 6 times longer than the (10, 0) unit cell, a large supercell of 17.1 Å along the nanotube axis was chosen to allow separation between adjacent chromophores.

**Acknowledgement**

We thank Prof. Judith Burstyn and Rob Mcclain for access to spectroscopic facilities. PG acknowledges useful discussions with Prof. Arnold (University of Wisconsin-Madison). We acknowledge financial support from the Division of Materials Sciences and Engineering, Office of Basic Energy Science, U.S. Department of Energy under Award #ER46590. DJM acknowledges support from NSF award #1005462.

*Supporting Information Available:* UV-Vis spectra of the three hybrids before and after washing with MeOH, Raman Spectra of the three chromophores and the hybrids, and PL spectrum of the three hybrids. This material is available free of charge *via* the Internet at http://pubs.acs.org.

**REFERENCES AND NOTES**
(1) Komarov, V. M.; Kayushin, L. P. *Studia Biophysica* **1975**, *52*, 107-140.
(2) Darwin, C. *The Power of Movement in Plants*; Murray: London, 1880.
(3) Schoenlein, R.; Peteanu, L.; Mathies, R.; Shank, C. *Science* **1991**, *254*, 412-415.
(4) Crescitelli, F. *Progress in Retinal Research* **1991**, *11*, 1-32.
(5) Barrett, C. J.; Mamiya, J.-i.; Yager, K. G.; Ikeda, T. *Soft Matter* **2007**, *3*, 1249-1261.
(6) Das, B.; Abe, S. *The Journal of Physical Chemistry B* **2006**, *110*, 4247-4255.
(7) Wen, Y.; Yi, W.; Meng, L.; Feng, M.; Jiang, G.; Yuan, W.; Zhang, Y.; Gao, H.; Jiang, L.; Song, Y. *The Journal of Physical Chemistry B* **2005**, *109*, 14465-14468.
(8) Yasuda, S.; Nakamura, T.; Matsumoto, M.; Shigekawa, H. *Journal of the American Chemical Society* **2003**, *125*, 16430-16433.
(9) Rotkin, S.; Zharov, I. *International Journal of Nanoscience* **2002**, 347-355.
(10) Simmons, J. M.; In, I.; Campbell, V. E.; Mark, T. J.; Leonard, F.; Gopalan, P.; Eriksson, M. A. *Physical Review Letters* **2007**, *98*, -.
(11) Ohno, Y.; Kishimoto, S.; Mizutani, T. *Jpn. J. Appl. Phys.* **2005**, *44*, 1592-1595.

(12) Freitag, M.; Martin, Y.; Misewich, J. A.; Martel, R.; Avouris, P. *Nano Letters* **2003**, *3*, 1067-1071.
(13) Zhao, Y.-L.; Stoddart, J. F. *Accounts of Chemical Research* **2009**, *42*, 1161-1171.
(14) Zhou, X.; Zifer, T.; Wong, B. M.; Krafcik, K. L.; LeÌ•onard, F. o.; Vance, A. L. *Nano Letters* **2009**, *9*, 1028-1033.
(15) Feng, Y. Y.; Zhang, X. Q.; Ding, X. S.; Feng, W. *Carbon* **2010**, *48*, 3091-3096.
(16) Bahr, J. L.; Mickelson, E. T.; Bronikowski, M. J.; Smalley, R. E.; Tour, J. M. *Chemical Communications* **2001**, 193-194.
(17) Kim, H.; Colavita, P. E.; Paoprasert, P.; Gopalan, P.; Kuech, T. F.; Hamers, R. J. *Surface Science* **2008**, *602*, 2382-2388.
(18) Paoprasert, P.; Spalenka, J. W.; Peterson, D. L.; Ruther, R. E.; Hamers, R. J.; Evans, P. G.; Gopalan, P. *Journal of Materials Chemistry* **2010**, *20*, 2651-2658.
(19) Reilman, R. F.; Msezane, A.; Manson, S. T. *Journal of Electron Spectroscopy and Related Phenomena* **1976**, *8*, 389-394.
(20) Laibinis, P. E.; Bain, C. D.; Whitesides, G. M. *Journal of Physical Chemistry* **1991**, *95*, 7017-7021.
(21) Zhao, Y.; Truhlar, D. G. *Accounts of Chemical Research* **2008**, *41*, 157-167.
(22) Chen, R. J.; Zhang, Y.; Wang, D.; Dai, H. *Journal of the American Chemical Society* **2001**, *123*, 3838-3839.
(23) D'Souza, F.; Chitta, R.; Sandanayaka, A. S. D.; Subbaiyan, N. K.; D'Souza, L.; Araki, Y.; Ito, O. *Journal of the American Chemical Society* **2007**, *129*, 15865-15871.
(24) Lu, J.; Nagase, S.; Zhang, X.; Wang, D.; Ni, M.; Maeda, Y.; Wakahara, T.; Nakahodo, T.; Tsuchiya, T.; Akasaka, T.; Gao, Z.; Yu, D.; Ye, H.; Mei, W. N.; Zhou, Y. *Journal of the American Chemical Society* **2006**, *128*, 5114-5118.
(25) Rao, A. M.; Richter, E.; Bandow, S.; Chase, B.; Eklund, P. C.; Williams, K. A.; Fang, S.; Subbaswamy, K. R.; Menon, M.; Thess, A.; Smalley, R. E.; Dresselhaus, G.; Dresselhaus, M. S. *Science* **1997**, *275*, 187-191.
(26) Moonoosawmy, K. R.; Kruse, P. *Journal of the American Chemical Society* **2008**, *130*, 13417-13424.
(27) Fleming, O. S.; Stepanek, F.; Kazarian, S. G. *Macromolecular Chemistry and Physics* **2005**, *206*, 1077-1083.
(28) Marino, I. G.; Bersani, D.; Lottici, P. P. *Optical Materials* **2001**, *15*, 279-284.
(29) Biswas, N.; Umapathy, S. *J. Chem. Phys.* **2003**, *118*, 5526–5536.
(30) Rao, A. M.; Bandow, S.; Richter, E.; Eklund, P. C. *Thin Solid Films* **1998**, *331*, 141-147.
(31) Okpalugo, T. I. T.; Papakonstantinou, P.; Murphy, H.; McLaughlin, J.; Brown, N. M. D. *Carbon* **2005**, *43*, 153-161.
(32) Wong, B. M. *Journal of Computational Chemistry* **2009**, *30*, 51-56.
(33) Zhao, Y.; Truhlar, D. G. *Journal of the American Chemical Society* **2007**, *129*, 8440-8442.
(34) Shen, Y. R. *Annual Review of Physical Chemistry* **1989**, *40*, 327-350.
(35) Simpson, G. J.; Rowlen, K. L. *Accounts of Chemical Research* **2000**, *33*, 781-789.
(36) Zyss, J.; Chemla, D. S. *Nanlinear Optical Properties of Organic Molecules and Crystal*; Academic Press: Orlando, FL, 1987; Vol. 1.
(37) Lalama, S. J.; Garito, A. F. *Physical Review A* **1979**, *20*, 1179-1194.

(38) Loucifsaibi, R.; Nakatani, K.; Delaire, J. A.; Dumont, M.; Sekkat, Z. *Chemistry of Materials* **1993**, *5*, 229-236.
(39) Ahmad, A.; Kern, K.; Balasubramanian, K. *ChemPhysChem* **2009**, *10*, 905-909.
(40) Ahmad, A.; Kurkina, T.; Kern, K.; Balasubramanian, K. *ChemPhysChem* **2009**, *10*, 2251-2255.
(41) Zhu, Z.; Yang, R. H.; You, M. X.; Zhang, X. L.; Wu, Y. R.; Tan, W. H. *Anal Bioanal Chem* **2010**, *396*, 73-83.
(42) Dettlaff-Weglikowska, U.; SkÃ¡kalovÃ¡, V.; Graupner, R.; Jhang, S. H.; Kim, B. H.; Lee, H. J.; Ley, L.; Park, Y. W.; Berber, S.; TomÃ¡nek, D.; Roth, S. *Journal of the American Chemical Society* **2005**, *127*, 5125-5131.
(43) Kim, K. K.; Bae, J. J.; Park, H. K.; Kim, S. M.; Geng, H.-Z.; Park, K. A.; Shin, H.-J.; Yoon, S.-M.; Benayad, A.; Choi, J.-Y.; Lee, Y. H. *Journal of the American Chemical Society* **2008**, *130*, 12757-12761.
(44) Zhou, W.; Vavro, J.; Nemes, N. M.; Fischer, J. E.; Borondics, F.; Kamar; aacute; s, K.; Tanner, D. B. *Physical Review B* **2005**, *71*, 205423.

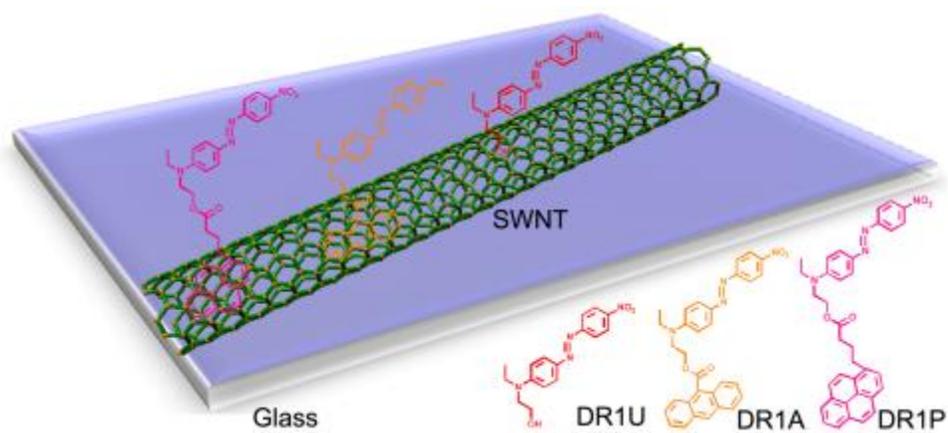

**Scheme 1.** Structure of unmodified DR1 (DR1U), DR1 with anthracene (DR1A), DR1 with pyrene (DR1P) tethers and their SWNT hybrids.

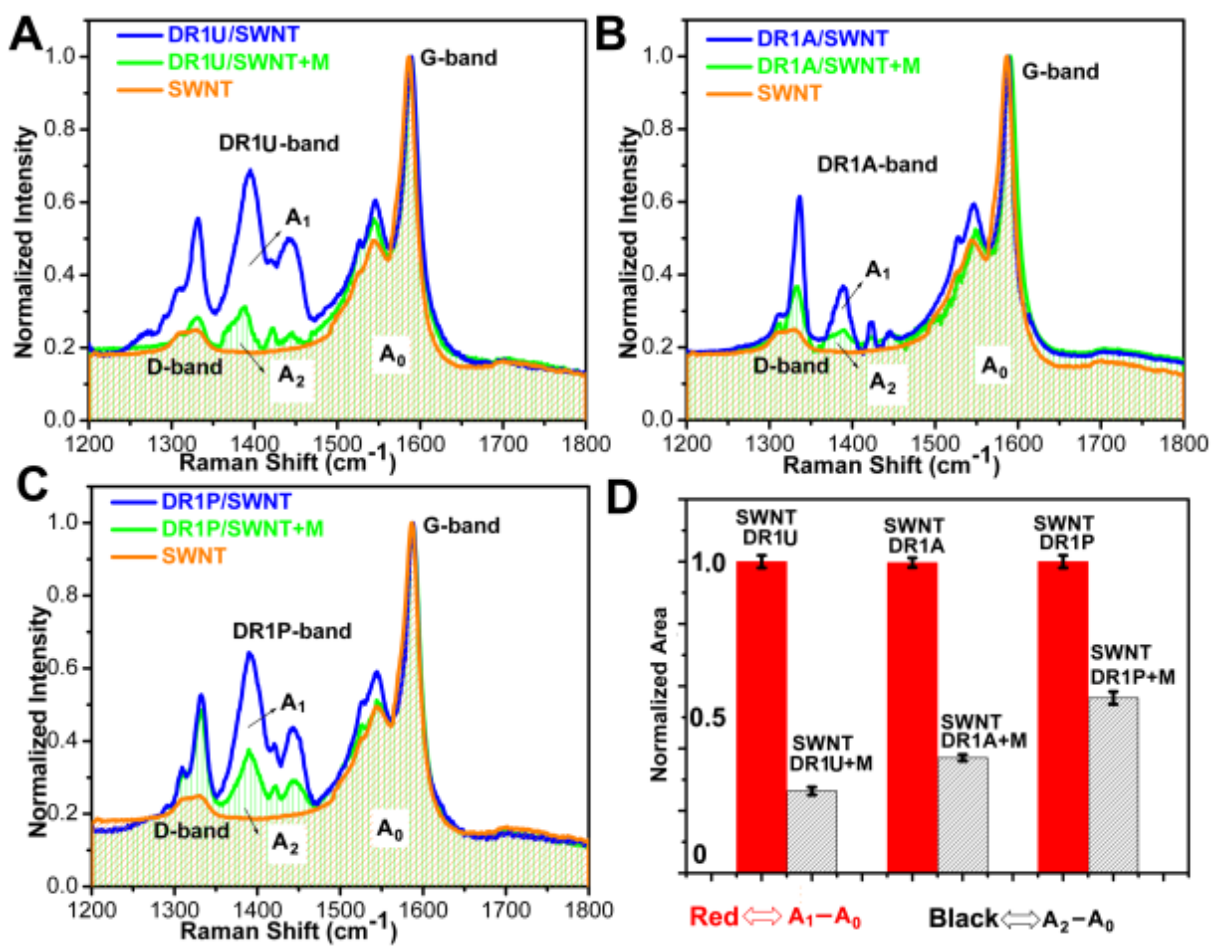

**Figure 1.** Raman spectra of SWNTs (orange), chromophore-SWNT hybrids (blue) and the same hybrids after washing with methanol (green). From plots (a) and (b), it is clear that methanol washing removes a significant fraction of the DR1U and DR1A. Plot (c) shows results for DR1P where methanol is less effective in removing DR1P from the SWNTs. Plot (d) gives a quantitative assessment of the normalized efficiency of binding for DR1U, DR1A and DR1P. Normalization was done with respect to the peak at 1590 cm$^{-1}$ (see supporting information for more details). Red bars correspond to the absorption intensity from the bound chromophores [difference in the peak areas represented $A_1$ (absorption intensity after binding the chromophores to SWNTs) and $A_0$ (absorption intensity from the bare nanotubes)]; and the grey bars correspond to the absorption intensity from the bound chromophores remaining after methanol wash [the difference in the peak areas represented $A_2$ (absorption intensity after washing the chromophore bound SWNTs with methanol) and $A_0$ (absorption intensity from the bare nanotubes)].

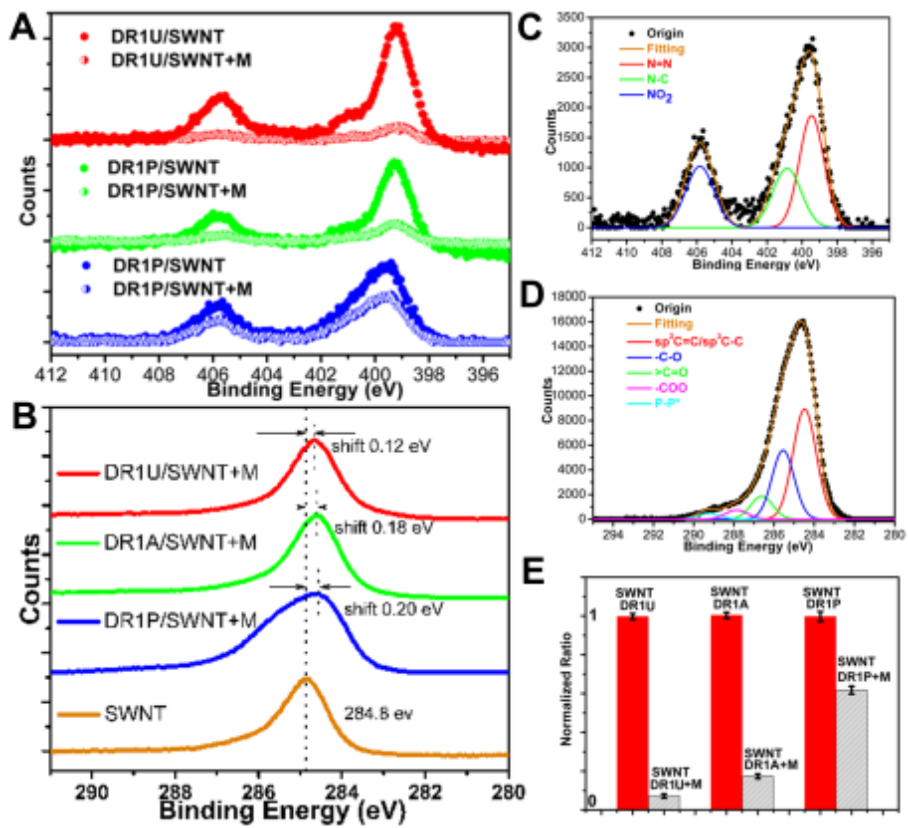

**Figure 2**. XPS spectra of chromophore-SWNT films showing: A) XPS core level N (1s) spectra for the three nanohybrid films as cast and after methanol washing, B) XPS core level spectra of C (1s) in pristine and doped SWCNT films after methanol washing, C) N (1s) spectra of the DR1P/SWNT hybrid with the fitting, D) C (1s) spectra of DR1P/SWNT hybrid with the fitting, and (E) normalized surface coverage of chromophores on SWNTs calculated by comparison of the area under the N (1s) peak with the area under the C (1s) peak.

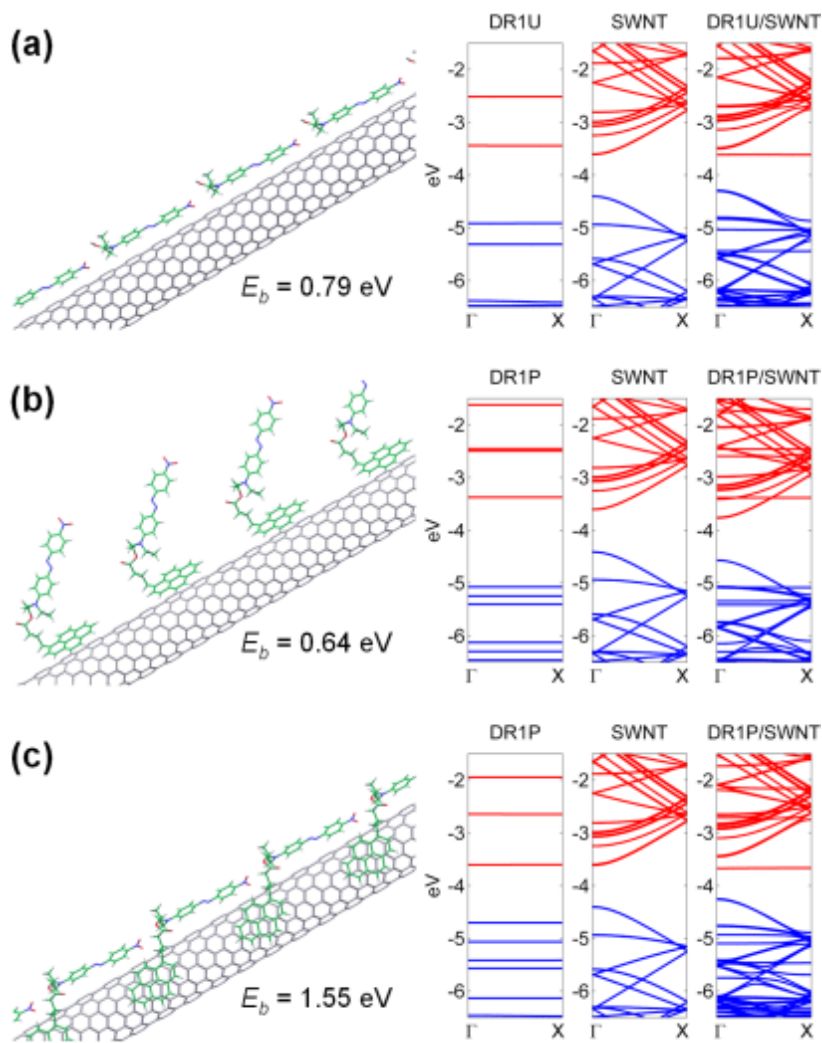

**Figure 3**. Optimized geometries and electronic band structures of three nanohybrids from DFT calculations using the M06-L/6-31G(d,p) level of theory. The DFT binding energies indicate that DR1P is more strongly adsorbed to the SWNT in the parallel geometry [(c) Eb = 1.55 eV] than in the perpendicular geometry [(b) Eb = 0.64 eV]. If a perpendicular arrangement is assumed, the DR1P is more weakly bound than (unmodified) DR1U [(a) Eb = 0.64 eV]. The band structures of the nanohybrids (last column) show some hybridization between the chromophore with the SWNT, particularly in energy regions near the X symmetry point. In all three cases, the fundamental band gap of the SWNT is nearly unchanged

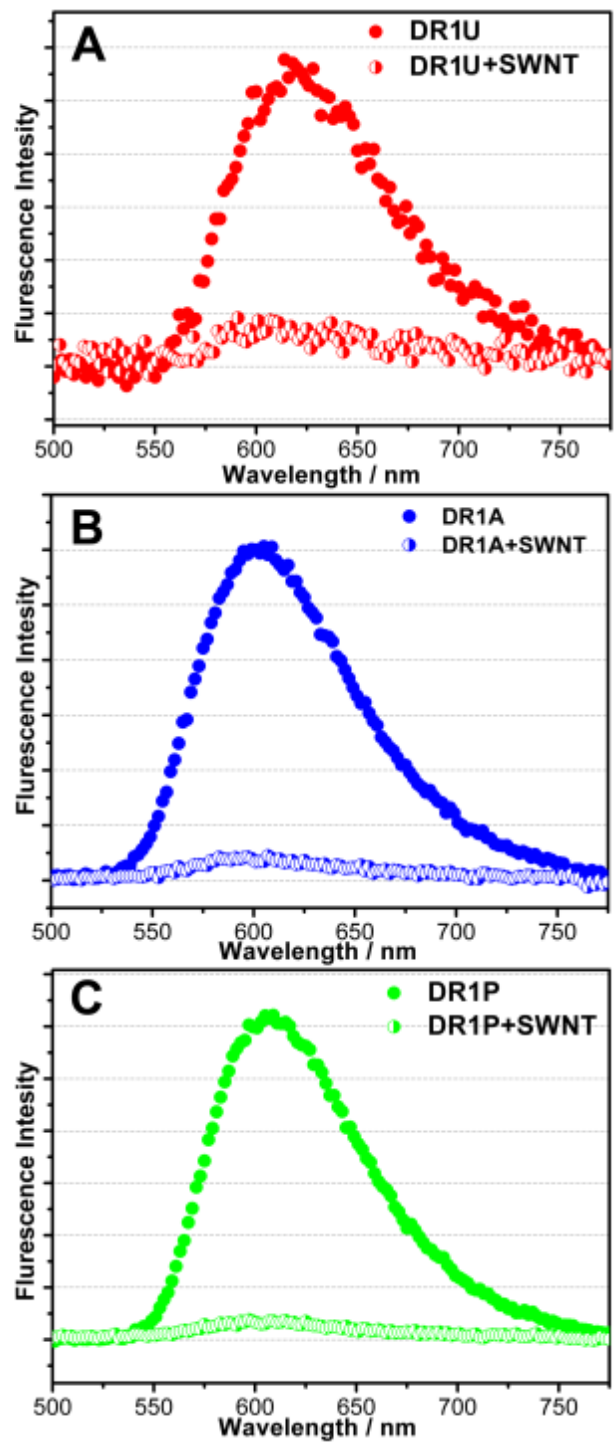

**Figure 4.** Emission spectra showing quenching of the fluorescence upon binding for 3 cases: (a) 0.01mM DR1U and 0.01mM DR1U with SWNT, (b) 0.01mM DR1A and 0.01mM DR1A with SWNT, (c) 0.01mM DR1P and 0.01mM DR1P with SWNT. Excitation wavelength = 490 nm.

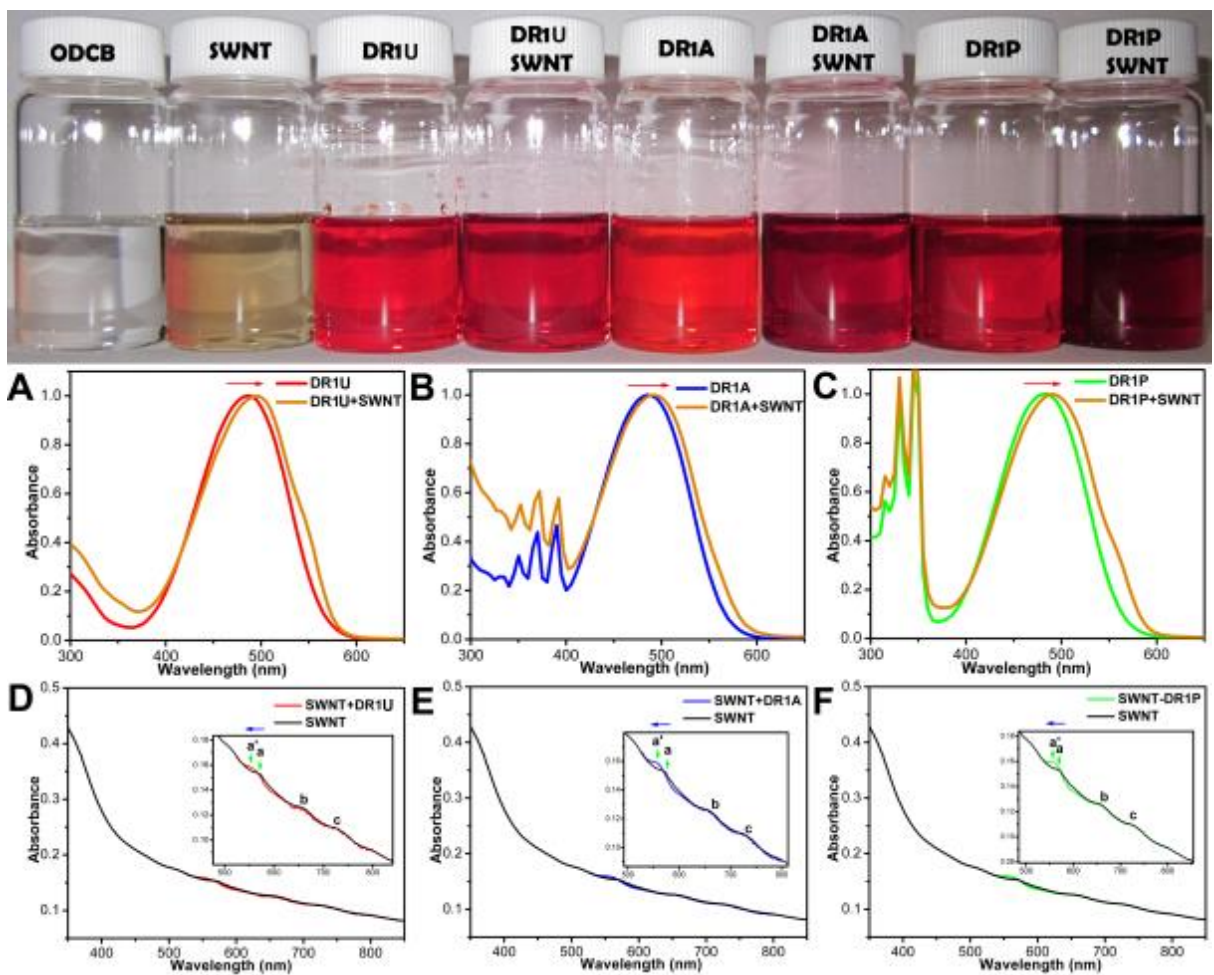

**Figure 5.** Absorption spectrum of (a) DR1U and DR1U/SWNT hybrid, (b) DR1A and DR1A/SWNT hybrid, (c) DR1P and DR1P/SWNT hybrid. UV-vis-NIR absorption spectrum of (d) SWNT and DR1U/ SWNT hybrid, (e) SWNT with DR1A/ SWNT hybrid, (f) SWNT and with DR1P/ SWNT. The arrows in the insets to panels (d-f) indicate shifts in the absorbance spectrum arising from specific chrial or diameter subsets of carbon nanotubes. The top row of labeled photographs show the prepared solutions.

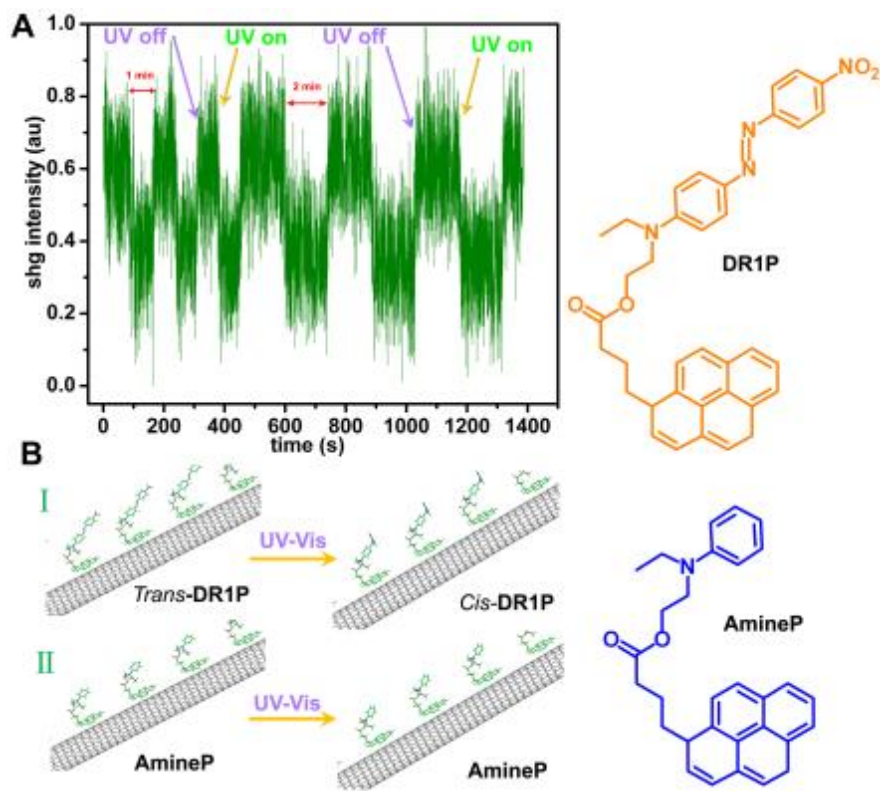

**Figure 6**. A) Second harmonic generation (SHG) signal from DR1P/SWNT film subject to cyclic UV irradiation in 1 and 2 minute intervals. For example, UV is on for 100 s < t < 160 s, off for 160 s < t < 220 s. At t = 460 s, the period is increased to 2 minutes. B) Schematic of the isomerization of DR1P on SWNTs and AmineP as control on SWNTs.

# Supporting Information:

# Spectroscopic properties of nanotube-chromophore hybrids

Changshui Huang[†], Randy K. Wang[‡], Bryan M. Wong[⊥], David McGee[£], François Léonard[⊥], Yun Jun Kim[†], Kirsten F. Johnson[‡], Padma Gopalan[†,§,*], and Mark A. Eriksson[‡,*]

*† Department of Materials Science & Engineering, University of Wisconsin-Madison, 1150 University Ave., Madison, Wisconsin 53706*
*‡ Department of Physics, University of Wisconsin-Madison, 1150 University Ave., Madison, Wisconsin 53706*
*§ Department of Chemistry, University of Wisconsin-Madison, 1150 University Ave., Madison, Wisconsin 53706*
*⊥ Sandia National Laboratories, Livermore, California 94551*
*£ Department of Physics, Drew University, 36 Madison Ave., Madison, New Jersey 07940*

E-mail: pgopalan@wisc.edu, maeriksson@wisc.edu

**Normalization of the XPS:** In the XPS spectra of the nanohybrids, the N (1s) peak has contributions from the azo, nitro, and the amine groups on the chromophore; the C (1s) peak has contributions from both the carbon atoms of chromophore as well as the SWNTs. Given that the ratio of the N:C atom in the chromophore is a constant value, we calculated the amount of C arising solely from the SWNTs.

In order to get a sense for the relative binding strengths, nanohybrids films were fabricated by drop-casting on to a glass substrate (blue line in Figure 1s) and washed with methanol to remove any weakly bound molecules (green curve in Figure 1s) and the UV-Vis spectrum were compared. The washing protocol is described in the experimental section. In the nanohybrid spectra (blue) the characteristic broad peaks of DR1U, anthracene, and pyrene appear at ~ 490 nm, 340 ~ 400 nm, and 300 ~ 380 nm respectively.

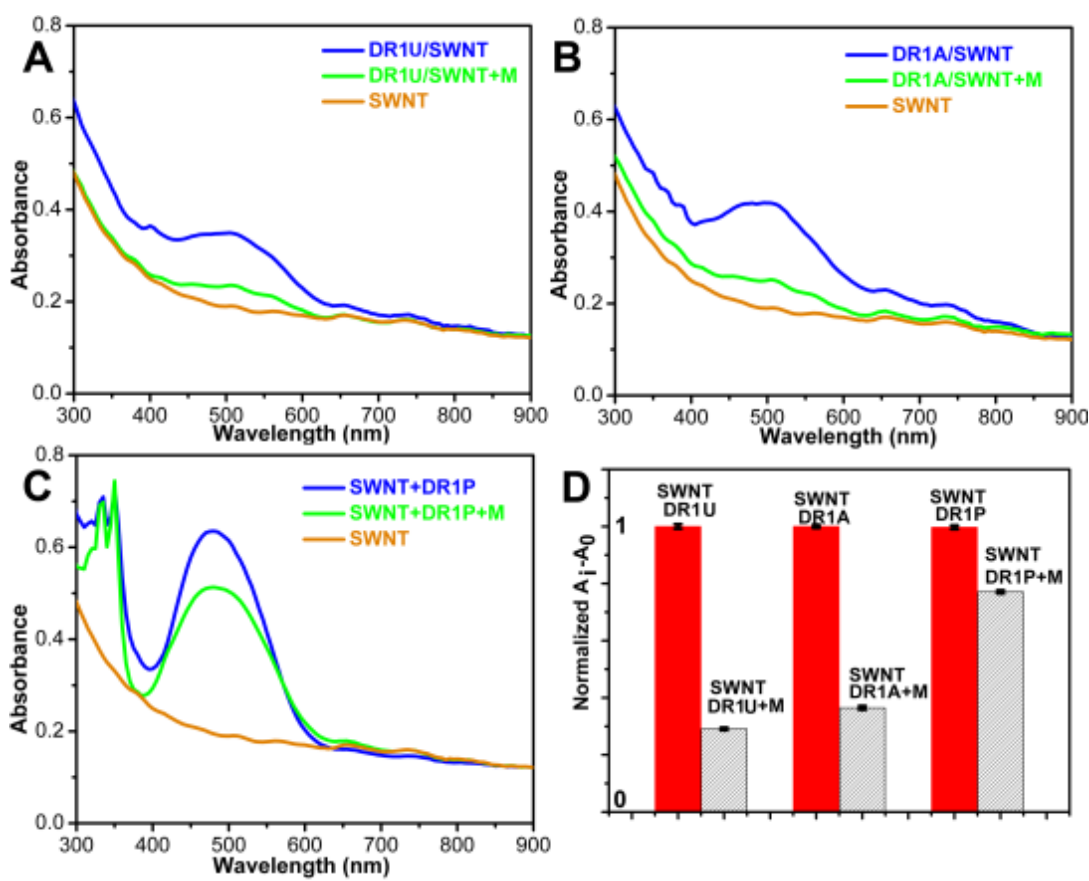

**Figure 1s.** UV-Vis spectra of SWNTs (red) and chromophore/SWNTs nanohybrids: (a) DR1U/SWNTs, (b) DR1A/SWNTs, and (c) DR1P/SWNTs film on glass.

In all cases, the as-cast chromophore-SWNTs nanohybrids films exhibited stronger absorbance than the methanol washed films. According to Beer's law, the absorbance is directly proportional to the concentration of the molecules. Hence the absorbance in blue curves in Figure 1s is an additive effect of both weakly and strongly bound molecules on SWNTs. However, after washing with methanol, the absorption (green curve) is predominantly from the chromophores bound to SWNTs. Comparing the normalized adsorption intensities (see experimental section for details) before and after methanol washing, it is clear that DR1U and DR1A are more weakly bound compared to DR1P, hence confirming that pyrene is probably a better tether than anthracene for non-covalent functionalization.

**Normalization of the Raman spectra:** The Raman spectra (excitation wavelength = 532 nm) in Fig. 1 are normalized to the peak corresponding to the G-band (1590 cm$^{-1}$). Three areas within the limits 1200-1800 cm$^{-1}$ are marked in figure 1: $A_0$ corresponds to the area under the absorption spectra resulting only from the nanotubes; $A_1$ corresponds to the total area under the absorption spectra resulting from the chromophore/nanotube hybrid, and $A_2$ corresponds to the total area under the absorption spectra resulting from the chromophore/nanotube hybrid following methanol rinse. For normalization of Raman spectra (figure1d): Red bars correspond to the absorption intensity from the bound chromophore [$A_1 - A_0$] which is set to be 1; and the grey bars correspond to the absorption intensity from the bound chromophores remaining after methanol rinse [$A_2-A_0$ / $A_1- A_0$]. The UV-vis spectrum was normalized similarly. For XPS, the ratio of nitrogen to nanotube carbon atoms in SWNT-chromophore hybrid film was set to be 1.

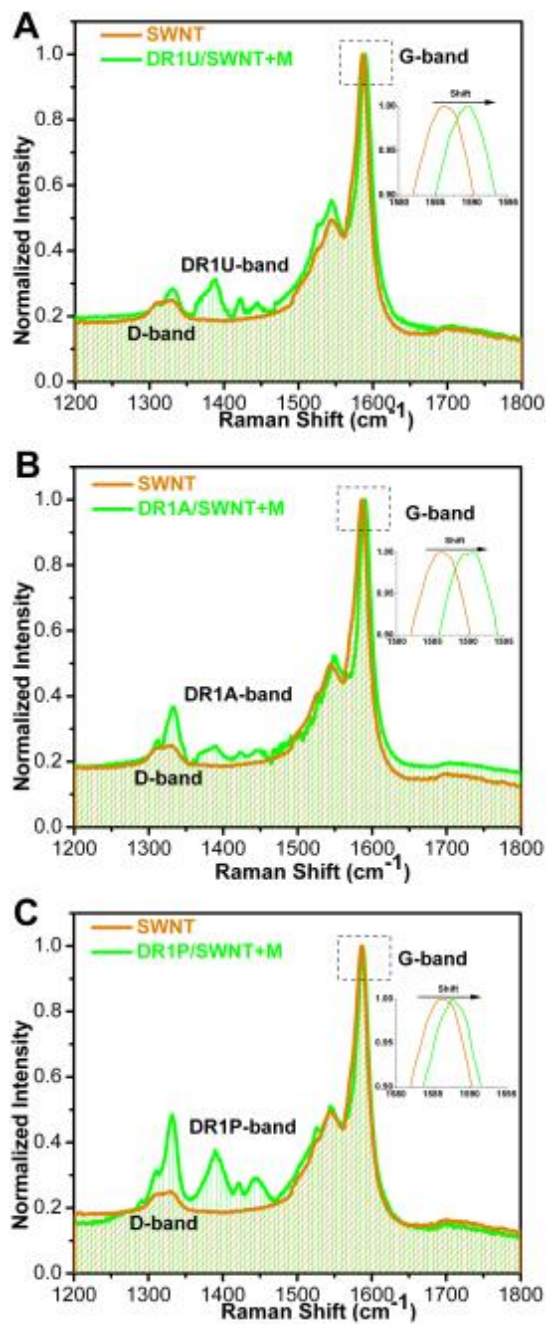

**Figure 2S**. Raman spectrum of chromophore-SWNTs films after methanol washing show up-shift compared to the Raman spectrum of pristine SWCNT film.

In our Raman data, the G-band position of pristine SWNTs is at 1586 cm$^{-1}$, measured for a network of SWNTs Introducing DR1U to SWNT caused the G-band to shift to 1589 cm$^{-1}$ after washing, which is about a 3 cm$^{-1}$ shift DR1A deposition caused the G-band to shift to 1590 cm$^{-1}$, while DR1P/SWNT shows a peak at 1588 cm$^{-1}$.

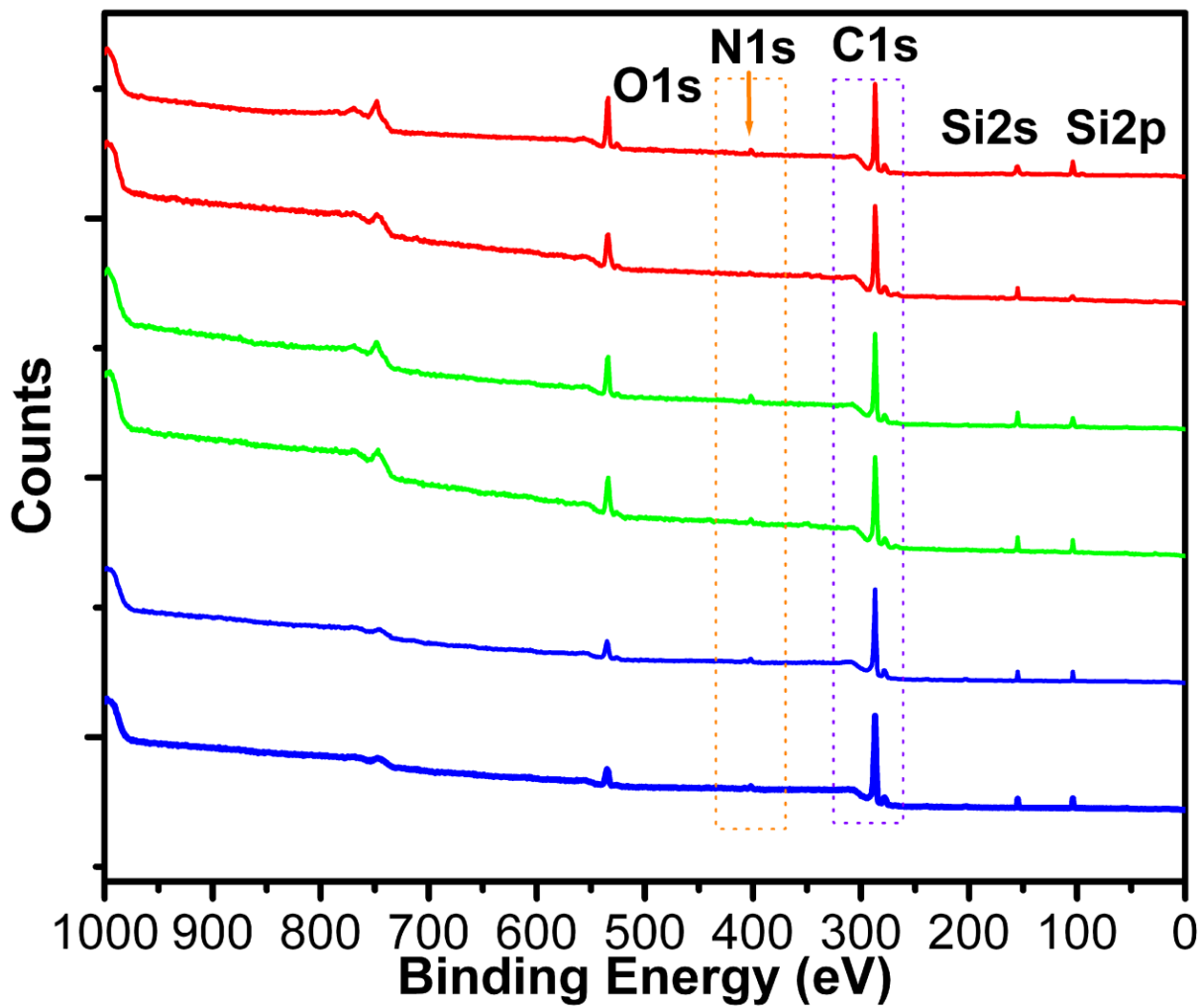

**Figure 3S**. XPS spectra survey of chromophore/SWNTs films. DR1U/SWNT before and after washing by methanol(red), DR1A/SWNT before and after washing by methanol (green), DR1P/SWNT before and after washing by methanol (blue).